\documentclass[aps,twocolumn,floatfix,showpacs,showkeys]{revtex4}
\usepackage{graphicx}
\usepackage{epsfig}
\input{psfig.sty}

\newcommand{\bastar}{\begin{eqnarray*}}
\newcommand{\eastar}{\end{eqnarray*}}
\newskip\humongous \humongous=0pt plus 1000pt minus 1000pt

\newif\ifdtup

\relax
\newcommand{\be}{\begin{equation}}
\newcommand{\ee}{\end{equation}}
\newcommand{\bea}{\begin{eqnarray}}
\newcommand{\eea}{\end{eqnarray}}
\newcommand{\pro}{\partial}
\newcommand{\n}{\hat n}
\newcommand{\oneg}{\displaystyle\frac{1}{g}}
\newcommand{\D}{{\hat D}}

\newcommand{\W}{{\vec W}}
\newcommand{\vW}{{\vec W}}

\newcommand{\A}{{\vec A}}

\newcommand{\astW}{\stackrel{\ast}{W}}

\newcommand{\dfrac}{\displaystyle\frac}
\newcommand{\ba}{\begin{array}}
\newcommand{\ea}{\end{array}}
\newcommand{\nn}{\nonumber}

\newcommand{\valpha}{{\vec \alpha}}

\begin{document}
\title{Knot soliton in Weinberg-Salam model}

\author{B. A. Fayzullaev $ ^{a,}$ } 
%\email{bfayzullaev@nuuz.uzsci.net}
%\affiliation{Theoretical Physics Department,
%Uzbekistan National University, Tashkent 700174, Uzbekistan}
\author{M. M. Musakhanov $ ^{a,b}$}
%\email{musakhanov@pusan.ac.kr}
%\affiliation{Department of Physics,
%Pusan National University, 609-735 Busan, Republic of Korea}
%\affiliation{Theoretical Physics Department, Uzbekistan National
%University, Tashkent 700174, Uzbekistan}
\author{D. G. Pak $ ^{c}$}
%\email{dmipak@phya.snu.ac.kr}
%\affiliation{ Center for Theoretical Physics,
%Seoul National University, Seoul 151-742, Korea}
\author{M. Siddikov $ ^{a,d}$ }
%\email{Marat.Siddikov@TP2.Ruhr-Uni-Bochum.de}

\affiliation
{  \vskip 4 mm  $ ^a $ Theoretical Physics Department,
Uzbekistan National University, Tashkent 700174, Uzbekistan \\
 $ ^b$ Department of Physics,
Pusan National University, 609-735 Busan, Republic of Korea \\
$ ^c $ Center for Theoretical Physics,
Seoul National University, Seoul 151-742, Korea \\
$ ^d $ 
Institut f\"ur Theoretische Physik II,
Ruhr-Universit\"at Bochum, D-44780 Bochum, Germany}
%\affiliation{Theoretical Physics Department, Uzbekistan National
%University, Tashkent 700174, Uzbekistan}

\begin{abstract}

We study numerically the topological knot solution  
suggested recently in the Weinberg-Salam model.
Applying the $SU(2)$ gauge invariant Abelian projection
we demonstrate that the restricted
part of the Weinberg-Salam Lagrangian containing
the interaction of the neutral boson with the
Higgs scalar can be reduced to the Ginzburg-Landau model with
the hidden $SU(2)$ symmetry.
The energy of the knot composed from the neutral boson
and Higgs field has been
evaluated by using the variational method with  
a modified Ward ansatz. The obtained numerical value 
is 39 Tev which provides the upper bound on the 
electroweak knot energy.
\end{abstract}

\pacs{12.15.-y, 14.80.-j, 11.27.+d}
\keywords{knot,standard model}
\maketitle

{\bf 1. Introduction}

It has been well known since 1980's that
knot-like topological solitons
corresponding to a non-trivial Hopf number
exist in some non-linear $\sigma$-models
with a scalar field $n^a, a=1,2,3$ \cite{vak,vlad,kundu}.
Last decade a wide class of such topological knot solitons
have been proposed in various physical models:
in the effective theory
of quantum chromodynamics (QCD) \cite{fadd,batt},
in two-gap superconductivity models
\cite{scond}, and in two-component Bose-Einstein condensate
theory \cite{bec,batt2,chobec}. Recently it was suggested that
knot solitons could exist even in high energy physics,
namely, in
the Weinberg-Salam model \cite{chows}.

In the present paper we consider numerically the properties of
the knot soliton made of a neutral $Z$ boson with a dressing
Higgs field
in the Weinberg-Salam model.
We apply the Ward ansatz \cite{ward}
which provided a high accuracy for the energy of the Faddeev-Niemi
knot  and results in, as we will show below,
a reasonable value 39 Tev for the energy
of the electroweak knot.
Even though the obtained energy value seems to be
out of practical detection with the present experimental facilities,
the electroweak knot may have an impact on cosmological problems
related to dark matter and baryon asymmetry of the universe
\cite{rubak,shap,tur}. In particular, it was suggested that
the formation of primordial hypermagnetic knots and their further
decay could result in the baryon asymmetry \cite{shap,giov}.
We suppose that the existence of the topological knot in the standard model
can lead to an alternative mechanism of the origin of the baryon asymmetry
in the universe, like the baryogenesis
scenario based on electroweak strings \cite{brand}.

{\bf 2. Gauge invariant Abelian projection}

First we consider the $SU(2)$ Yang-Mills theory with a complex Higgs scalar
doublet $\phi$ as a simple model of Weinberg-Salam theory.
We will extract from the Lagrangian of the Weinberg-Salam model the part
containing the interaction of the neutral boson
with the Higgs scalar using the gauge invariant Abelian projection.
By this way we keep explicitly the full $SU(2)$ gauge invariance
and at the same time we preserve the topological structure of the
original group $SU(2)$ in the Abelianized part of the 
gauge connection.
This approach
provides the gauge invariant definition of the Hopf number
in terms of the Abelian potential and gives rise to the description
of the neutral boson interacting with the Higgs scalar in the framework of
a simple Ginzburg-Landau model.

We will follow the approach proposed in \cite{cho1, chows}.
Let us start from the gauge invariant
Abelian projection which is given by the
following decomposition of the full $SU(2)$ connection
\cite{cho1}
\bea \label{decomposition}
& \vec{A}_\mu =A_\mu \n - \oneg \n\times\pro_\mu\n+\vW_\mu\nonumber
         = \hat A_\mu + \vW_\mu, \nn\\
& (A_\mu = \n\cdot \vec A_\mu,~ \n^2 =1,~ \hat{n}\cdot\vec{W}_\mu=0).
\eea
The restricted potential $\hat A_\mu$ possesses transformation
properties of the full $SU(2)$ connection and leads to
the desired Abelianization of the theory.
The vector part of the restricted potential
$\vec C_\mu \equiv - \oneg \n\times\pro_\mu\n$
is defined in terms of the triplet scalar
field $\n$ whose configuration is classified
by Hopf invariant, i.e. by a homotopy group
$\pi_3(S^2)$. The scalar field $\n$
is covariantly constant
\bea
\D_\mu \n = \pro_\mu \n + g {\hat A}_\mu \times \n = 0
\eea
and represents the topological degrees of freedom
in the theory.
The $SU(2)$ gauge transformations  can be written
as follows
\bea
\delta \n = - \vec \alpha \times \n  \,,\,\,\,\,
\delta \A_\mu = \oneg  D_\mu \vec \alpha,
\eea
or
\bea
&&\delta A_\mu = \oneg \n \cdot \pro_\mu \valpha,\,\,\,\
\delta \hat A_\mu = \oneg \D_\mu \valpha  ,  \nn \\
&&\hspace{1.2cm}\delta \W_\mu = - \valpha \times \W_\mu  .
\eea

In a local orthonormal frame $(\n_1, \n_2, \n_3 \equiv \n)$ of the
internal $SU(2)$ space one can rewrite
the decomposition (\ref{decomposition}) as
\bea
&&\vec{A}_\mu =A_\mu \n + (\omega_\mu^{(1,2)}
 + \vW_\mu^{(1,2)}) \n_{(1,2)} , \label{decomp1}
\eea
where $\omega_\mu^{(1,2)}, \,{\vec W}_\mu^{(1,2)} $
are local components of the vectors ${\vec C}_\mu, {\vec W}_\mu$ 
correspondingly.
By appropriate gauge transformation one can
pass to the magnetic gauge \cite{cho1} in which
the vector potential
$\vec C_\mu $
converts into the Abelian magnetic potential $C_\mu$
\bea
&& \vec A_\mu = (A_\mu + C_\mu) \n + \vW_\mu \label{decomp2}, \nn \\
&& C_\mu = -\oneg \n_1 \pro_\mu \n_2.
\eea

With the decomposition (\ref{decomposition}) one 
can factorize the Abelian potential part and the off-diagonal
components in an explicit gauge invariant manner 
\bea
\vec{F}_{\mu\nu}&=&(F_{\mu \nu}+H_{\mu\nu}) \hat n  \nn \\
&&+ \D _\mu \vW_\nu -
\D_\nu \vW_\mu + g\vW_\mu \times \vW_\nu,
\eea
where
\bea
&&F_{\mu\nu}= \pro_\mu A_\nu - \pro_\nu A_\mu, \nn \\
&&H_{\mu\nu} = -\dfrac{1}{g} \epsilon^{abc}
        \n^a \pro_\mu \n^b \pro_\nu \n^c =
\pro_\mu C_\nu-\pro_\nu C_\mu .
\eea
So that the Lagrangian can be rewritten in the form
\bea
&&{\cal L} = -\dfrac{1}{4} \vec F^2_{\mu \nu } \nn\\
&&=-\dfrac{1}{4}
(F_{\mu\nu}+H_{\mu\nu})^2 -\dfrac{1}{4}(\D_\mu\vW_\nu-\D_\nu\vW_\mu)^2-
\nn \\
&&\dfrac{g}{2} (F_{\mu\nu}+H_{\mu \nu}) \cdot (\vW_\mu \times \vW_\nu)
-\dfrac{g^2}{4} (\vW_\mu \times \vW_\nu)^2 .
\eea

In complex notation for the charged vector boson
\bea
&W_\mu = \dfrac{1}{\sqrt{2}} ( W^1_\mu + i W^2_\mu ),
\eea
one can express the Lagrangian explicitly in terms of
the Abelian potential 
$A_\mu + C_\mu$ and the charged vector field $W_\mu$,
\bea \label{lagrangian}
&&{\cal L}=-\dfrac{1}{4} (F_{\mu\nu}+ H_{\mu\nu})^2
-\dfrac{1}{2}|{D}_\mu{W}_\nu-{D}_\nu{W}_\mu|^2  \nn \\
&&+ig( F_{\mu\nu}+ H_{\mu\nu}) \astW_\mu W_\nu
-\dfrac{g^2}{2} \Big[(\astW_\mu W_\mu)^2 \nn \\
&& -(\astW_\mu)^2 (W_\nu)^2 \Big] 
= -\dfrac{1}{4}(F_{\mu\nu} + H_{\mu\nu} + W_{\mu\nu})^2 \nn \\
&&-\dfrac{1}{2}|{D}_\mu{W}_\nu-{D}_\nu{W}_\mu|^2,
\end{eqnarray}
where
\bea
&& W_{\mu\nu} = - i g ( \astW_\mu W_\nu - \astW_\nu W_\mu ), \nn\\
&&{D}_\mu{W}_\nu = (\partial_\mu + ig (A_\mu+C_\mu)) W_\nu.  \nonumber
\eea
It should be stressed,
that the Lagrangian is invariant not only under the original
$SU(2)$ group transformations but
under the action of the dual $\tilde U(1)$ group as well
\bea
&& \delta_{SU(2)} (A_\mu + C_\mu) = 0, \nn \\
&& \delta_{\tilde U(1)} (A_\mu + C_\mu) = \pro_\mu \tilde \lambda.
\eea
Due to this the Hopf number can be defined as
a Chern-Simons index of $SU(2)$ gauge theory.
For the case $A_\mu=0$ employed later on we have the standard formula
for the Hopf number
\cite{cho1}
\bea
N_H = \dfrac{g^2}{32 \pi^2} \int d^3 x \epsilon_{ijk} C_i H_{jk} .
      \label{eqHopf}
\eea
Notice,
the Abelian symmetry ${\tilde U}(1)$ has been appeared
in QCD as a dual magnetic symmetry
\cite{cho1} which is an essential ingredient 
of the dual Meissner effect leading to the monopole mechanism
of confinement.

With these preliminaries let us start from the
bosonic sector of the Weinberg-Salam Lagrangian
\bea
&& {\cal L} =
-\dfrac{1}{4} ({\vec F}_{\mu\nu})^2-\dfrac{1}{4} (G_{\mu\nu})^2
  - |(\pro_\mu -\dfrac{ig}{2} \vec \sigma \cdot \vec A_\mu \nn \\
&& -ig' B_\mu)
{\tilde \phi} |^2 -
\dfrac{\lambda}{2} ({\tilde \phi}^\dagger {\tilde \phi} -
 \dfrac{v^2}{2})^2, \nn \\
&&G_{\mu\nu}=\pro_\mu B_\nu - \pro_\nu B_\mu,
\eea
where $B_\mu$ is the Abelian potential corresponding to the
weak hypercharge group $U_Y(1)$.

We consider the restricted case with the vanishing charged vector
boson $W_\mu=0$ and electromagnetic field $A_\mu=0$
while keeping only the Higgs field ${\tilde \phi}$
and the neutral $Z$ boson in the Lagrangian.
We will identify the classical part of $Z$ boson
containing the whole topological content of the knot
with a hypermagnetic potential $C_\mu$
\cite{chows}.
It is convenient to express the fields $\n , {\tilde \phi}$ in terms
of the real components
$\rho, \alpha$ and $SU(2)$ complex doublet $\xi$
which parametrizes the complex projective space
$CP^1$
\bea
&&\n^a = \xi^+ {\vec \sigma}^a \xi, \nn \\
&& {\tilde \phi} = \dfrac{1}{\sqrt 2} \xi \rho e^{i \alpha}\equiv
      \dfrac{1}{\sqrt 2} \xi \phi, \nn \\
&& \xi^+ \xi = 1.  \label{eqn}
\eea
Since the hypermagnetic potential $C_\mu$ is
the dual potential one can express it locally as follows
\bea
C_\mu = \dfrac{4i}{\bar g} \xi^+ \pro_\mu \xi. \label{eqC}
\eea
With this we obtain the final expression for the
Lagrangian for our purpose to
study the static knot type solution
\bea
&&{\cal L} =-\dfrac{1}{4} H_{\mu \nu}^2-|(\pro_\mu+
i \dfrac{\bar g}{2} C_\mu) \phi |^2
 - \dfrac{\lambda }{2} (|\phi|^2-\rho_0^2)^2 . \nn \\
&& \label{GL}
\eea

Formally, the expression (\ref{GL}) can be easily recognized as a well-known
Ginsburg-Landau Lagrangian.
But one should stress that
the original non-Abelian structure of the Lagrangian (\ref{GL}) is
hidden and encoded in the topological properties of the
hypermagnetic potential
$C_\mu$ given by (\ref{eqC}). By this the non-trivial homotopy group
$\pi_3(CP^1)$ ensures the existence of the topological knot
solution classified by
Hopf number $N_H$.

{\bf 3. Knot soliton}

The idea of a non-topological string in the electroweak theory
was proposed long time ago
\cite{nambu}. Later the twisted electroweak strings
with a baryon number had been introduced in
\cite{vach}, and the existence of the knot
in the Weinberg-Salam model was suggested
recently
\cite{chows}.
We will find the static topological knot solution
by minimization of the energy functional
corresponding to the Lagrangian (\ref{GL}) ($i=1,2,3$)
\bea
&& E =\int d^3x {\cal E} = \int d^3x \Big [
 \dfrac{1}{4} H_{ij}^2 +
\dfrac{1}{2}(\pro \rho)^2 \nn \\
&&  + \dfrac{{\bar g}^2}{8}
\rho^2 (C_i+\frac{2}{\bar g} \pro_i \alpha)^2 +
 \dfrac{\lambda}{8}(\rho^2- v^2)^2 \Big ].
\eea

Let us introduce the dimensionless variables
$ r \rightarrow r m_W, \,\,
 k \equiv  2 \sqrt \lambda /g = m_H/m_W ,\,\,
C_i\rightarrow C_i g/m_W,\,\, \rho \rightarrow \rho/v $
with rescaling $\alpha \rightarrow \alpha \cos \theta_W$
and 
 $m_W =80.22 $ Gev , $m_H=200 $ Gev, $\, \sin^2
\theta_W=0.2325$. 
One can  rewrite the energy functional
in terms of the dimensionless variables as follows
\bea
&&  E = \dfrac{m_W}{g^2}  \int d^3 x \Big [\dfrac{1}{2} {\vec H}^2 +
 2 (\pro \rho)^2  \nn \\
&&+  \dfrac{1}{2 \cos^2 \theta_W} \rho^2
(C_i+2 \pro_i \alpha)^2 +\dfrac{k^2}{2} (\rho^2-1)^2 \Big ],
\eea
where $(\vec H)_i= \epsilon_{ijk} \pro_j C_k$
 is the hypermagnetic field strength of the
neutral boson.

To find out appropriate trial functions
we start from the original Ward ansatz \cite{ward}.
The stereographic projection of
$\hat n$ is given by a complex field
\bea
Z = \dfrac{n_1 + i n_2}{1+ n_3} .
\eea
The Ward ansatz \cite{ward} with a radial trial function $f(r)$
\bea
&&Z = \dfrac{x + iy}{z - i f(r)} , \nn \\
&& f(r) = a ( r-b)(r^2 + c r + d)
\eea
was proposed to describe the Faddeev-Niemi knot soliton
with a good accuracy for the energy value.
The non-trivial Hopf number $N_H=1$ is determined by the boundary conditions
of the function $f(r)$ at points $r=0 , \, \infty$.
Applying
this ansatz with the trial function $f(r)$
and the definitions (\ref{eqn}) one can
find the corresponding 
expression for the complex doublet $\xi$ 
(in spherical coordinates)
\bea
&&\xi= \left( \begin{array}{c}
          \sqrt {\frac{h^2(r)+\cos^2 \theta}{1+h^2(r)}} \\
    \frac{ \sin \theta}{\sqrt {1+h^2(r)}} e^{i \psi(r,\theta, \varphi)}
\end{array} \right), \nn \\
&& \tan \psi(r,\theta,\varphi) = \frac{\sin \varphi \cos \theta + h(r) \cos \varphi}
            {\cos \varphi \cos \theta -h(r) \sin \varphi} ,
\eea
where $h(r)=f(r)/r$.
With this one can write down the components of the 
magnetic potential 
\bea
 C_r &=& \dfrac{2 h' \sin^2 \theta \cos \theta}
    {(1+ h^2)(h^2 + \cos^2 \theta)} , \nn \\
C_\theta &=&  \dfrac{2 h \sin^3 \theta}
    {r (1+ h^2)(h^2 + \cos^2 \theta)} , \nn \\
 C_\varphi &=& \dfrac{2 \sin \theta}{r (1 + h^2)}.
\eea

One can use the gauge freedom to remove the
dependence of the phase factor
$\psi(r, \theta,\varphi)$ on the variables $(r, \theta)$.
By making an appropriate gauge transformation
\bea
&&C_i\rightarrow C_i -2  \nabla_i \arctan (\frac{h}
{\cos \theta }) , \nn \\
&& \psi(r,\theta,\varphi) \rightarrow \psi(r,\theta,\varphi)-\arctan (\frac{h}
{\cos \theta})
\eea
one results in
a simple parametrization for the phase factor
$\psi(r,\theta,\varphi) = \varphi$ and for the components
$C_r, C_\theta$ 
\bea
&&C_r=-\frac{2 h' \cos \theta}{1+h^2} , \nn \\
&&C_\theta=-\frac{2 h \sin \theta }{r(1+h^2)} . \label{eqCh}
\eea

The phase factor in the scalar field
$\phi = \rho e^{i\alpha}$
appears in the Lagrangian only in the combination  $C_i + 2\pro_i \alpha$.
We will adopt the following ansatz
for the phase factor $\alpha (r, \theta) = \beta (r) \cos \theta$
with the boundary conditions $\beta (0)= \beta (\infty)=0$.
The chosen form of the
ansatz preserves
the Hopf number calculated with the gauge transformed
potentials
\bea
&&C_r=-\frac{2  h' \cos \theta}{1+h^2} +2 \beta'(r)
            \cos \theta , \nn \\
&&C_\theta=-\frac{2 h \sin \theta}{r(1+h^2)} -
 \dfrac{2 \beta (r)}{r} \sin \theta , \nn \\
&& C_\varphi=\frac{2 \sin \theta }{r(1+h^2)} . \label{modward}
\eea
It turns out that the modified ansatz (\ref{modward})
provides lower value of the minimum
 of the energy functional. 
For the Higgs field $\rho(r, \theta)$
we impose the boundary conditions
\bea
\rho (0, \theta) = c, \,\,\,\, \rho(\infty, \theta) =1 ,
\eea
where $c$ is a number.
The minimization of the energy functional gives
the energy of the knot
$E_0=39$ Tev.  This energy value has been checked
with a class of various trial functions including
the angle dependence on $\theta$. We found that the 
minimum of the energy functional is reached 
for a spherically symmetric configuration of the Higgs field.
We expect that our modified Ward
ansatz (\ref{modward}) with a pure radial trial function
for the Higgs field $\rho(r)$ 
provides the main contribution
to the knot energy similar to
the case of the Faddeev-Niemi knot. 
The energy density plots in the plane $XY$ and $XZ$ are shown
in Figs. 1, 2 from which one can see that the
energy of the knot is localized along the torus.

The topological stability of the knot is provided
by non-vanishing hypermagnetic fluxes along and around the torus.
The total hypermagnetic flux around the axis $0Z$ is
$\Phi_\varphi = 2 \pi $.
The total hypermagnetic flux through the inner disk of the torus
($r<r_0=1.64,\theta=\pi/2 $) is $\Phi_z=2 \pi$.
By this we have the linking of two fluxes which provides the Hopf number
$Q_H=1$ and the topological stability. Notice, that the lines of the
hypermagnetic vector field $\vec H$ form in fact a twisted torus.
\begin{figure}[t]
\includegraphics[scale=0.72]{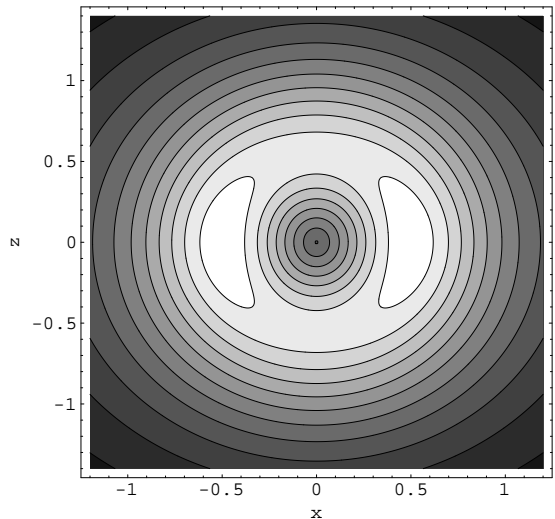}
\caption{The contour plot of the energy density of the knot
in the plane $y=0$. The maximal value of the energy density
is ${\cal E}_{max}=7.42 $ , the step size
between the contours is $0.39$ (in units of $v/g=0.407$ Tev/$m^3$).}
\label{Fig.1}
\end{figure}

One should stress, since we use a variational
method with a constrained ansatz
we can not rather obtain a more detailed structure
of the energy density distribution of the knot.
By plotting the energy density one can describe only
qualitatively the inner structure of the knot and
estimate roughly the effective size of the knot.
The energy density has a maximum along the circle
$r_0=0.50, \theta = \pi/2$
which corresponds to the
radius of the knot
$1.45\times 10^{-18}m$.
From the other hand we can define the radial size of the knot
as a radius of the circle at which the hypermagnetic component $H_\theta$
changes its sign. This gives the radius $r_0=4.1 \times 10^{-18 }m$
which is near the value $3.5 \times 10^{-18}m$
estimated in \cite{chows}.
Notice, that the 
Ward ansatz gives exactly the Dirac type magnetic
flux quantization
with the minimal non-zero 
flux $\Phi = 2 \pi$ due to the semi-simple structure of
the electroweak gauge group $SU(2) \times U_Y(1)$
and the non-compactness of the electric
charge group $U_e(1)$.
\begin{figure}[t]
\includegraphics[scale=0.7]{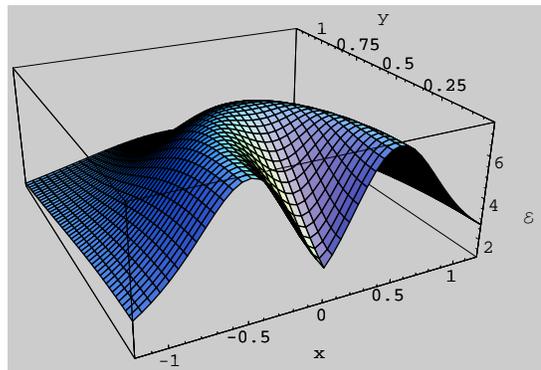}
\caption{ The energy density ${\cal E} (x,y,0)$ of the knot
in the plane $z=0$ }
\label{Fig.2}
\end{figure}
For a model with a simple gauge group
$SU(2)$ we would have Wu-Yang type flux quantization
with a flux quantum proportional to $4 \pi $. Such situation
has been realized, for instance,
in a two-component Bose-Einstein condensate system
\cite{chobec}. The energy density and 
hypermagnetic field components
$H_\theta, H_\varphi$ in the plane $\theta=\pi/2$, the
Higgs field $\rho(r)$ and the phase radial function $\beta(r)$
 are shown in Fig. 3.
 
\begin{figure}[t]
\includegraphics[scale=0.8]{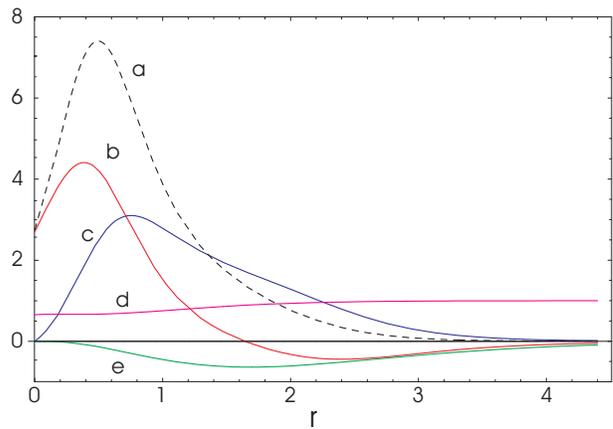}
\caption{ (a) the energy density ${\cal E} (r, \pi/2)$ 
(in unit of $v/g$);
(b) the hypermagnetic field $H_\theta (r,\pi/2)$; 
(c) the hypermagnetic field
$H_\varphi (r,\pi/2)$, (d) the Higgs field $\rho (r)$;
(e) the phase radial function $\beta (r) $ (in dimensionless units).}
\label{Fig.3}
\end{figure}

Since the energy of the knot is much greater than the 
energy of vector $W, Z$ bosons an important 
question arises whether the knot solution is stable
against the interaction with 
other fields, especially with the charged $W$ boson
and electromagnetic field. The only potentially dangerous 
part in the Weinberg-Salam Lagrangian which can cause
the instability is represented by the so called 
anomalous magnetic moment interaction term
$ ig H_{\mu\nu} \astW_\mu W_\nu$ in (\ref{lagrangian}).
This term leads to a severe vacuum instability problem
in QCD via generating a tachyonic mode \cite{niel}.

To analyse the stability of the knot soliton
we consider small fluctuations of $W$
field in knot background. The possible unstable modes 
are determined by the negative eigenvalues of the operator
$K$ defined by means of the second variational derivative 
of the Lagrangian with respect to infinitesimal
variations $\delta W_\mu$. In a suitable background gauge 
$(\pro_\mu +i g \cos \theta_W C_\mu) W_\mu=0$
one can find the operator $K$ as follows 
(in dimensionless variables)
\bea
&&K_{\mu \nu} = -\dfrac{m_W}{g^2} \Big [
 \eta_{\mu \nu} ( D_\rho D_\rho - \rho^2 )+
               i \frac{e}{g'} Z_{\mu \nu} 
                       \Big ],\nn \\
&& Z_{\mu \nu} = \pro_\mu C_\nu-\pro_\nu C_\mu,
\eea
where, $D_\mu = \pro_\mu + i \frac{e}{g'} C_\mu$.
One can factorize the eigenvalues of the matrix part with 
Lorentzian indices in the operator $K$ similarly to the case 
of derivation of functional determinants corresponding
to the one-loop effective action in $SU(2)$ QCD 
(details of derivation for general magnetic and electric 
background are given in \cite{choqcd})
\bea
&& K^{\pm} \simeq   -D_i D_i + \rho^2 \pm 2 \frac{e}{g'} H,  
\eea
where $H \equiv |\vec H|$.
With $C_\mu $ defined by (\ref{modward}) one can write down 
the eigenfunction equation for the operator $K^\pm$
\bea
&& \Big [-\dfrac{1}{r^2} \pro_r (r^2 \pro_r )-\dfrac{1}{r^2 \sin \theta}
\pro_\theta (\sin \theta \pro_\theta) -\frac{2ie}{g'} C_i \pro_i - \nn \\
&& \frac{ie}{g'} \pro_i C_i +\frac{e^2}{g'^2} C_i^2 +
\rho^2 \pm \frac{2e}{g'} H \Big ]\Psi(r, \theta)
 = \Lambda \Psi (r,\theta)  \label{scr}
\eea
The equation represents a Schrodinger type problem
for a quantum mechanical particle moving in a potential well. 
To estimate the lowest eigenvalue corresponding to the ground state
one can neglect the terms $2 i C_i \pro_i $ and  $ i \pro_i C_i $
which do not contribute to the ground state in the
leading order approximation due to the axial 
symmetry and linear dependency of the knot solution $C_i$ 
on angle functions.
The negative eigenvalues can arise only from the equation with
the lower negative sign in front of the last term
in l.h.s of (\ref{scr}). With this,
substituting the numerical knot solution into the
equation and solving it by the variational method
one can find the lowest eigenvalue which is found to be a 
positive number $\Lambda_0 = + 0.31\,$. 
The scalar field $\rho$ plays a role of the mass regulator
parameter which is close to the mass of $W$ boson (see Fig.3).
One can find numerically that negative modes would appear
if $\rho$ were replaced  with a constant averaged value 
less than $0.73$.
Surprisingly, despite on the facts that the knot energy is much 
higher than the energy of $W$ boson, and our solution is obtained 
with a simple modified Ward ansatz, nevertheless,
no any negative eigenvalues appear. 
The weak dependency on the knot energy is provided 
by the presence of two competetive terms in (\ref{scr}), 
$ \dfrac{e^2}{g'^2} C_i^2 $ and $- \dfrac{2e}{g'} H$,
which contribute opposite ways to the quantum mechanical potential
in the Schrodinger equation. We have checked also that
unstable modes do not appear under small deviation of the profile 
functions $f, \rho, \alpha$ from the numerical solution. 
This analysis gives us a hope that a rigorous exact solution for the 
knot will be stable as well.

In conclusion, we performed numerical study of 
the topological knot soliton in the Weinberg-Salam
model suggested in \cite{chows}. The obtained
numerical value of the knot energy
is 39 Tev which is higher
than the earlier
estimate 21 Tev \cite{chows}.
The last estimate
was based on the equation for the knot energy
obtained in the non-linear $\sigma$-model \cite{vak}
which represents actually the low energy bound. Since we
elaborate the energy of the knot with the variational
method, our numerical result provides the upper bound
which, we believe, is close to the real value for the energy
of the electroweak knot.
The knot energy value even being much higher
than the scale $100 $ Gev of the standard model, it is still
less than the scale of the temperature
$T\simeq 80 $ Tev for the right-electron equilibration
state considered in estimation of the baryon asymmetry 
\cite{camp,oliv}. This might be an indication that
the topological electroweak knot could play an important
role in cosmology
at the stage of formation of large scale structures
along with other proposed mechanisms.

{\bf Acknowledgements}

One of authors (D.G.P.) thanks Y. M. Cho
for useful discussions and helpful comments. This work was partially
supported by INTAS grant 2000-110.

\end{document}